\documentclass[onecolumn,prb]{revtex4-2}
\usepackage{graphicx,psfrag,color}
\def\bc{\begin{center}}
\def\ec{\end{center}}

\def\beq{\begin{equation}}
\def\eeq{\end{equation}}

\def\d{\downarrow}
\def\u{\uparrow}

\def\bc{{\bf c}}

\def\bb{{\bar b}}

\begin{document}

\title{
Chiral Josephson effect in double layers:\\
the role of particle-hole duality
}

\author{Klaus Ziegler} 
\affiliation{
Institut f\"ur Physik, Universit\"at Augsburg, D-86135 Augsburg, Germany
}
\date{\today}

\begin{abstract}
The Josephson effect of inter-layer s-wave pairing in a double layer of two chiral metals is considered. 
We employ the duality relation between electron-electron and electron-hole double layers to discuss
the zero-energy eigenmodes at a domain wall and their coupling to the superfluid state. This is described in 
terms of the quasiparticle current  and the supercurrent. It turns out that the degeneracy of
the zero-energy eigenmodes is resolved by the coupling to the supercurrent. The duality relation
between the electron-electron and electron-hole double layers leads to the same current distribution
in both systems but to different zero-energy modes.
\end{abstract}

\maketitle

\section{Introduction}

One of the most fascinating observations in condensed matter physics is the pairing effect, leading to phenomena such as
superconductivity and superfluidity. Although the pairing effect is of quantum nature, its theoretical description in terms
of a macroscopic order parameter field is given by a classical (mean-field) theory. Excitations in the form of quasiparticles, 
on the other hand, are represented by the Bogoliubov de Gennes (BdG) equation that describes a quantum wave 
function~\cite{sigal22}. A related interesting
phenomenon is the Josephson effect~\cite{JOSEPHSON1962251} that originates in a coupling between the 
macroscopic superfluid or superconducting order parameter with the quasiparticle 
modes~\cite{PhysRevB.25.4515,FURUSAKI1990967,PhysRevLett.66.3056,
PhysRevB.67.184505,SPUNTARELLI2010111,2018RSPTA.37680140S}. 
It has been discussed for different systems, including 
electron-hole bilayers~\cite{PhysRevLett.93.266801,PhysRevB.106.L220503} 
and electron-hole double layers~\cite{1976JETP...44..389L,doi:10.1063/1.4831671,2212.11161} and
electron-hole double-bilayers~\cite{PhysRevLett.110.146803,2017NatPh..13..751L}.
This effect has the potential for the development of new technologies. For instance, it has been used to create and
manipulate qubits in quantum computational devices~\cite{RevModPhys.73.357,Kockum2019,2211.10852}.
An important aspect of the quasiparticles is their sensitivity to the underlying spatial structure in terms of
geometry and topology, in particular, for zero-energy modes.
The interplay of the Josephson effect with the topological properties of the quasiparticle modes was recently discussed for an
electron-electron double layer (EEDL)~\cite{PhysRevLett.128.157001} and for an electron-hole double layer 
(EHDL)~\cite{2212.11161} separately. The purpose of the present paper is study the connection of these
two systems through the particle-hole duality and how this affects the zero-energy eigenmodes, the
coupling of these modes to the superfluid and the distribution of currents. This analysis might be useful for
future studies of more complex, multi-band systems with electron-electron and electron-hole pairing.

\begin{figure}[t]
a)
\includegraphics[width=0.3\linewidth]{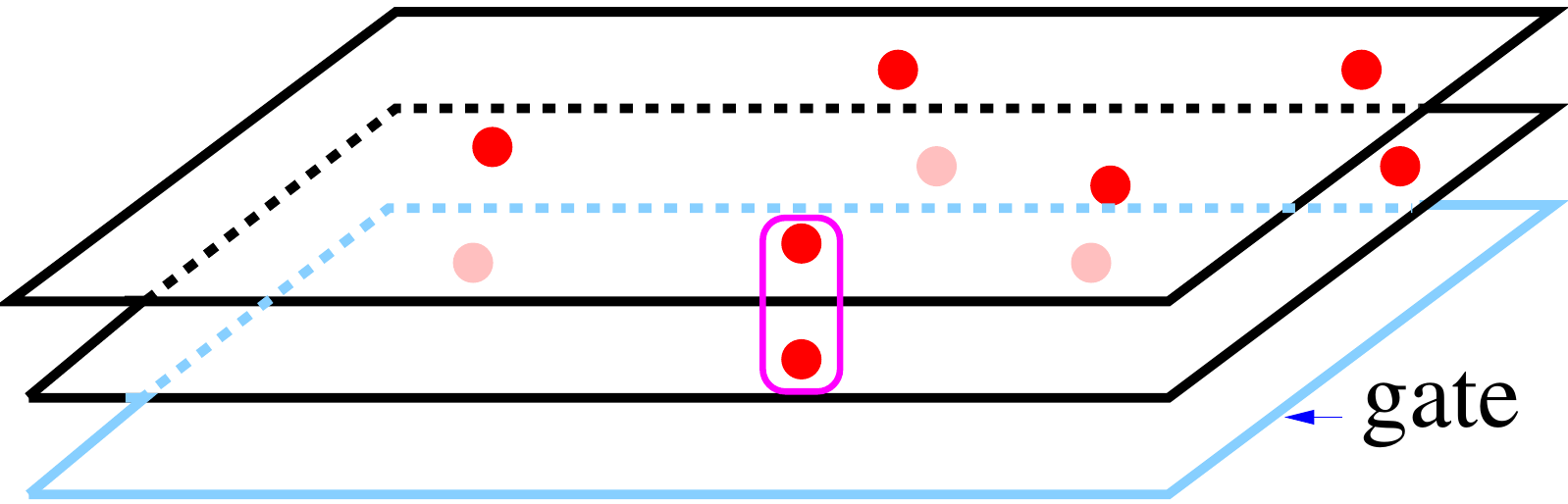}\\
\vskip0.5cm
b)
\includegraphics[width=0.3\linewidth]{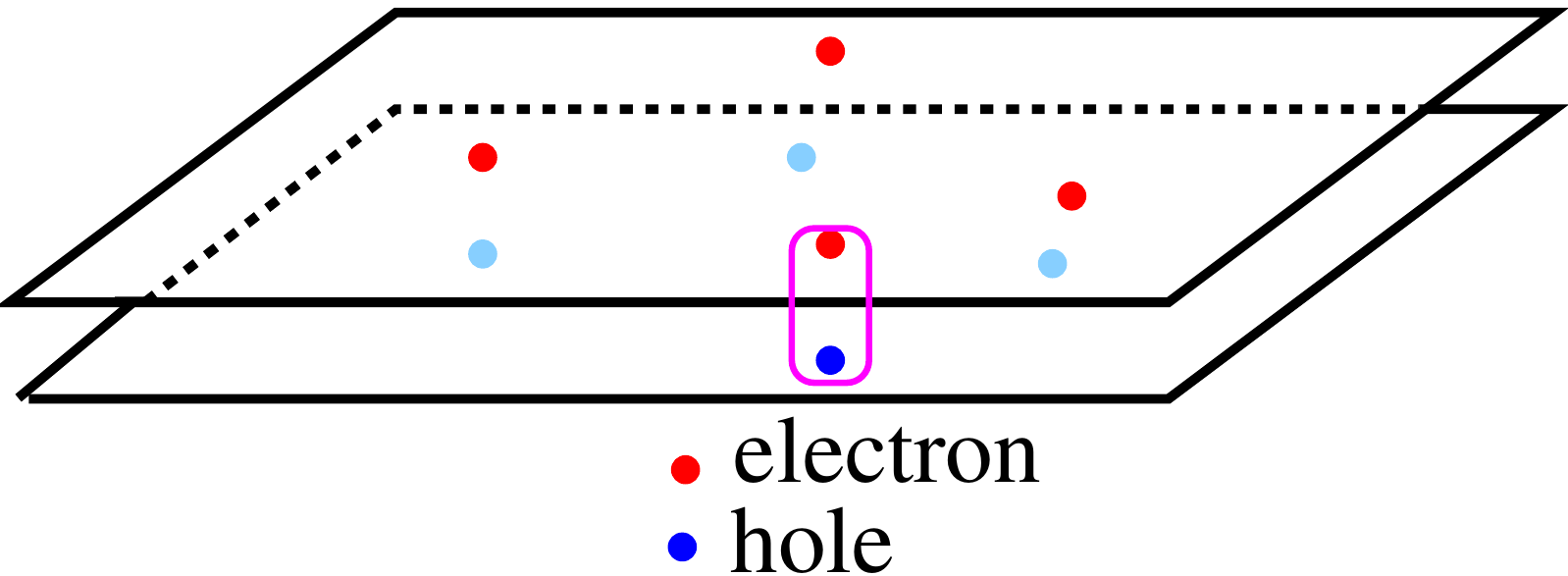}
\caption{An electron-electron double layer (a) and an electron-hole double layer (b) with inter-layer
pairing due to Coulomb interaction, where the schematic gate in a) is positively charged and guarantees charge neutrality.
Inter-layer tunneling is suppressed by a dielectric medium and
inter-layer pairing requires a small distance of the layers to make the Coulomb interaction sufficiently strong.
}
\label{fig:double_layer}
\end{figure}

%%%%%%%%%%%%%%%%%%%%%%%%%%%%%%%%%%%%%%%%%%%%%%%%
\section{Model: Bogoliubov de Gennes Equation/Hamiltonian}

The EEDL and the EHDL are dual to each other~\cite{2020PhRvR...2c3085S}. In the following we discuss these
two cases separately and compare the resulting Josephson currents in Sect. \ref{sect:discussion}. Both systems
are treated within a BCS-like mean-field approach. This leads to an order parameter $\Delta$ that
characterizes the superconducting state of the EEDL and the superfluid state of the EHDL. Excitations in the
form of quasiparticles are obtained form the corresponding BdG Hamiltonian, where the latter describes the
quantum fluctuations about the mean-field approximation. In the following discussion we consider the inter-layer
pairing but ignore the intra-layer pairing. This is a simplification which is plausible for a small distance
of the layers and due to screening inside the layers but has been debated in the 
literature~\cite{PhysRevB.45.415,pnas.2205845119}. 
Moreover, inter-layer tunneling is suppressed by a dielectric between the layers.

%%%%%%%%%%%%%%%%%%%%%%%%%%%%%%%%%%%%%%%%%%%%%%%%%%%
\subsection{Chiral electron-electron double layer} 

The EEDL comprises two electronic layers with a positively charged extra layer. The latter can either be an external
gate (as visualized in Fig. \ref{fig:double_layer}a) or is provided by the positive charges inside the metallic layers.
In both cases the entire system preserves charge neutrality. The electrons in the two layers repel each
other due to the Coulomb interaction. The geometric constraint enable the electrons at fixed density to 
form inter-layer Cooper pairs. This is formally supported by the duality transformation to the EHDL, in which the
electron-hole pairs are subject to an attractive Coulomb interaction. In other words, the formation of inter-layer 
electron-hole pairs in the EHDL~\cite{1976JETP...44..389L} is transformed into
to inter-layer electron-electron pairs by the duality transformation~\cite{2020PhRvR...2c3085S}. 
Then the related quasiparticles are described by the BdG Hamiltonian of two layers with opposite chiralities
reads~\cite{PhysRevLett.128.157001}
\begin{equation}
\label{hamiltonian0}
H^{}_{\rm EEDL}
=\pmatrix{
h_1\sigma_1+h_2\sigma_2 & \Delta\sigma_2 \cr
\Delta\sigma_2 & h_1\sigma_1-h_2\sigma_2\cr
}
,
\end{equation}
where $\sigma_j$ are Pauli matrices, $h_j$ are tight-binding hopping matrices and $\Delta$ is the pairing
order parameter. For the subsequent discussion we assume a honeycomb lattice for the underlying spatial 
structure of the tight-binding model, such that the quasiparticle Hamiltonian describes graphene-like materials.
Assuming translational invariance in $y$ direction,
the low-energy BdG Hamiltonian becomes with $h_1\sim i\hbar v_F\partial_x$, $h_2\sim \hbar v_F k_y$
\begin{equation}
\label{hamiltonian2}
H^{}_{\rm EEDL} 
\sim\pmatrix{
i\hbar v_F\partial_x\sigma_1+\hbar v_F k_y\sigma_2 & \Delta(x)\sigma_2 \cr
\Delta(x)\sigma_2 & i\hbar v_F\partial_x\sigma_1-\hbar v_F k_y\sigma_2\cr
}
,
\end{equation}
where $v_F$ is the Fermi velocity.

\begin{figure}[t]
\begin{center}
a)
\includegraphics[width=0.3\linewidth]{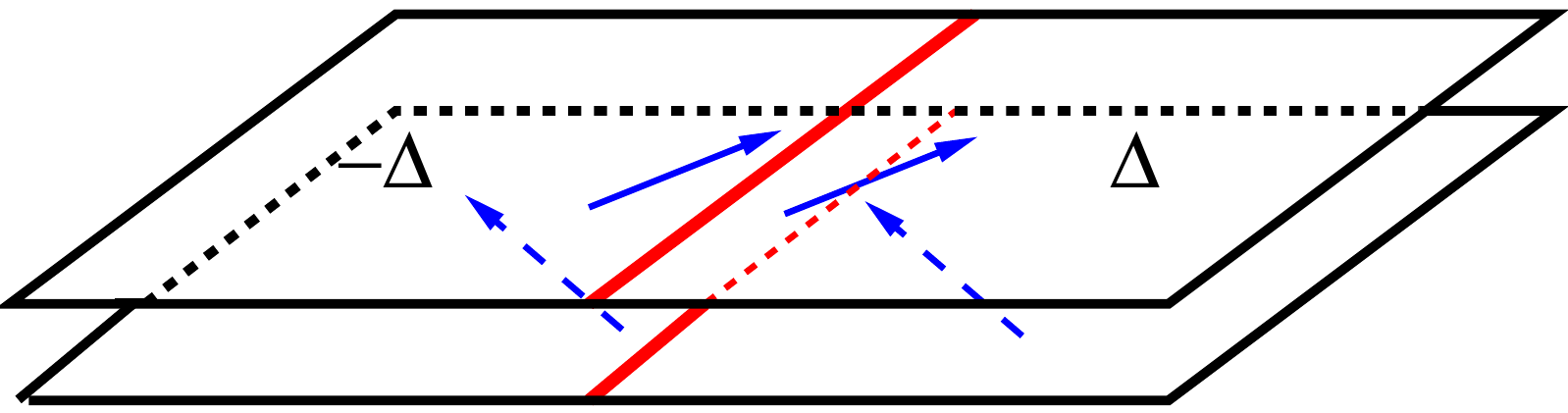}
b)
\includegraphics[width=0.3\linewidth]{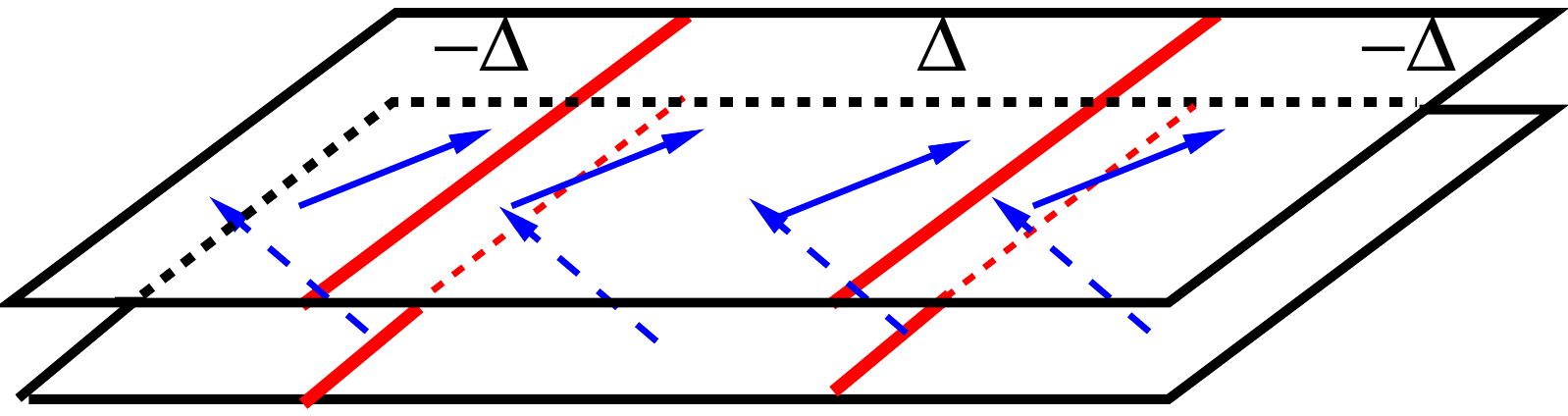}
\caption{
Electronic double layer with domain walls, which is given by a sign jump of the pairing order
parameter. The local currents (blue arrows) flow in the same (opposite) direction in the two layers 
parallel (perpendicular) to the domain wall. 
}
\label{fig:domain_wall}
\end{center}
\end{figure}

Now we consider a domain wall in $y$ direction at $x=0$, as sketched in
Fig.~\ref{fig:domain_wall}a: $\Delta(x)={\rm sgn}(x)|\Delta|$. The resulting eigenvalue problem can be solved. 
At zero energy there is an exceptional point for the Hamiltonian (\ref{hamiltonian2}), where the four-fold 
degeneracy coalesces to a two-dimensional eigenspace with two independent zero-energy modes 
(cf. App. \ref{app:exceptional}):
\beq
\label{zero_modes02}
\Psi_1=
\frac{1}{{\cal N}}\pmatrix{
1 \cr
0 \cr
1 \cr
0 \cr
}e^{-|\Delta||x|/\hbar v_F}
\ ,\ \ 
\Psi_2=\frac{1}{{\cal N}}\pmatrix{
0 \cr
1 \cr
0 \cr
-1 \cr
}e^{-|\Delta||x|/\hbar v_F}
\eeq
with the normalization ${\cal N}=\sqrt{2v_F\hbar/|\Delta|}$.
Any superposition of the two zero-energy modes $\Phi=a_1\Psi_1+a_2\Psi_2$ with complex coefficients
$a_j=|a_j|e^{i\varphi_j}$ (and normalization $|a_1|^2+|a_2|^2=1$) is also a zero-energy mode.
Thus, the zero-energy eigenmodes are complex in general and only real for a special choice of the coefficients.
These two modes are expressed separately for the top and for the bottom layer as
\beq
\label{zero_mode_EEDL}
\Phi_\u=\pmatrix{
a_1 \cr
a_2 \cr
}\frac{e^{-|\Delta||x|/\hbar v_F}}{{\cal N}}
\ ,\ \ 
\Phi_\d=\pmatrix{
a_1 \cr
-a_2 \cr
}\frac{e^{-|\Delta||x|/\hbar v_F}}{{\cal N}}
,
\label{layer_phi}
\eeq
which will be used for the calculation of the Josephson currents in Sect. \ref{sect:currents}.

%%%%%%%%%%%%%%%%%%%%%%%%%%%%%%%%%%%%%%%%%%%%%%
\subsection{Chiral electron-hole double layer}

The BdG Hamiltonian of the EHDL reads~\cite{2212.11161}
\begin{equation}
\label{Hamiltonian00}
H_{\rm EHDL}
=\pmatrix{
h_1\sigma_1+h_2\sigma_2 & \Delta\sigma_3 \cr
\Delta^*\sigma_3 & h_1\sigma_1+h_2\sigma_2\cr
}
,
\end{equation}
where the chirality of the two layers is the same now and the pairing order parameter appears with a Pauli matrix 
$\sigma_3$. This means 
that there is a coupling between the same metallic bands of the two layers with opposite sign though.
The BdG Hamiltonian is dual to the BdG Hamiltonian of the EEDL in Eq. (\ref{hamiltonian0}), and the 
duality transformation reads
\beq
\label{duality00}
H_{\rm EHDL}(i\Delta)=VH_{\rm EEDL}(\Delta)V
\ ,\ \
V=\pmatrix{
\sigma_0 & 0 \cr
0 & \sigma_1 \cr
}
,
\eeq
where the order parameter aquires a global imaginary unit.
This implies for the eigenvalue equation $H_{\rm EEDL}(\Delta)\Psi_E=E\Psi_E$
\beq
\label{duality01}
H_{\rm EHDL}(i\Delta)V\Psi_E=VH_{\rm EEDL}(\Delta)VV\Psi_E=EV\Psi_E
,
\eeq
i.e., $V\Psi_E$ is eigenmode of $H_{\rm EHDL}(\Delta)$ with eigenvalue $E$. Thus, the spectrum is invariant
under the duality transformation, whereas the zero-energy eigenmodes are not. In particular, from 
Eq. (\ref{zero_modes02}) for the domain wall $\Delta(x)=i{\rm sgn}(x) |\Delta|$
an exceptional point at zero energy and a two-dimensional eigenspace of zero-energy modes
\beq
\label{zero_modes02a}
V\Psi_1=
\frac{1}{{\cal N}}\pmatrix{
1 \cr
0 \cr
0 \cr
1 \cr
}e^{-|\Delta||x|/\hbar v_F}
\ ,\ \ 
V\Psi_2=\frac{1}{{\cal N}}\pmatrix{
0 \cr
1 \cr
-1 \cr
0 \cr
}e^{-|\Delta||x|/\hbar v_F}
.
\eeq
From these modes we can construct again the zero-energy modes of the individual layers as
\beq
\label{zero_mode_EHDL}
V\Phi=\pmatrix{
\Phi'_\u \cr
\Phi'_\d \cr
}
\ \ {\rm with}\ \ 
\Phi'_\u=\pmatrix{
a_1 \cr
a_2 \cr
}\frac{e^{-|\Delta||x|/\hbar v_F}}{{\cal N}}
\ ,\ \ 
\Phi'_\d=\pmatrix{
-a_2 \cr
a_1 \cr
}
\frac{e^{-|\Delta||x|/\hbar v_F}}{{\cal N}}
.
\eeq

%%%%%%%%%%%%%%%%%%%%%%%%%%%%%%%%%%%%%%%%%%%%%%%%
\section{Josephson currents}
\label{sect:currents}

With the help of the BdG equation we derive the continuity equation for the two layers separately 
as~\cite{PhysRevB.25.4515}
\beq
\label{cont00}
\partial_t\Phi_{\sigma}\cdot\Phi_{\sigma}+\partial_xI_{x\sigma}=0 \ \ (\sigma=\u,\d)
,
\eeq
where the total current $I=j+j^s$ is the sum of the quasiparticle current $j$ and the supercurrent $j^s$.
For the $y$ component we have $\partial_yI_{y\sigma}=0$ due to the uniform mode in the $y$
direction. The quasiparticle current operator of a BdG Hamiltonian $H_{\rm BdG}$ reads 
$j_{x}=\frac{i}{\hbar}[H_{\rm BdG},x]$.

The BdG equation of the EEDL yields for the continuity equation (\ref{cont00}) in the top layer
\beq
\label{cont_up}
\partial_t\Phi_\u\cdot\Phi_\u+\partial_xj_{x\u}
=i\frac{\Delta}{\hbar}\Psi^*_\d\sigma_2\Psi_\u-i\frac{\Delta^*}{\hbar}\Psi^*_\u\sigma_2\Psi_\d
=2{\rm sgn}(x)\frac{|\Delta|^2}{v_F\hbar^2}Re(a_1a_2^*)e^{-2|\Delta||x|/\hbar v_F}
\eeq
and in the bottom layer
\beq
\label{cont_down}
\partial_t\Phi_\d\cdot\Phi_\d+\partial_xj_{x\d}
=i\frac{\Delta^*}{\hbar}\Psi^*_\u\sigma_2\Psi_\d-i\frac{\Delta}{\hbar}\Psi^*_\d\sigma_2\Psi_\u
=-2{\rm sgn}(x)\frac{|\Delta|^2}{v_F\hbar^2}Re(a_1a_2^*)e^{-2|\Delta||x|/\hbar v_F}
.
\eeq
The expressions on the right-hand side of the equations are equal up to a minus sign.
The quasiparticle currents $j_{x\u}$,  $j_{x\d}$ are directly calculated from the commutator
$j_{x}=\frac{i}{\hbar}[H_{\rm EEDL},x]$, which gives
\beq
\label{qp_current_x}
j_{x\u}=-j_{x\d}
=-\frac{|\Delta|}{\hbar}Re(a_1a_2^*)e^{-2|\Delta||x|/\hbar v_F}
,
\eeq
such that we obtain from the continuity equations after integration along the $x$ direction 
the $x$-components of the supercurrents as
\beq
\label{super_current_x}
j^s_{x\u,\d}(x)=\mp\frac{ |\Delta|}{\hbar}Re(a_1a_2^*)(1-e^{-2|\Delta||x|/\hbar v_F})
.
\eeq
The supercurrent component $j^s_{x\u,\d}(x)$ vanishes at the domain wall $x=0$ and becomes
$\mp\frac{ |\Delta|}{\hbar}Re(a_1a_2^*)$ for $|x|\gg \hbar v_F/|\Delta|$.
Finally, from $j_{y}=\frac{i}{\hbar}[H_{\rm EEDL},y]$ we get the $y$-components of the quasiparticle currents
\beq
\label{qp_current_y}
j_{y\u}=j_{y\d}=-v_F\Phi_\d\cdot\sigma_2\Phi_\d=-\frac{|\Delta|}{\hbar}Im(a_1a_2^*)e^{-2|\Delta||x|/\hbar v_F}
.
\eeq
The corresponding continuity equation of the EHDL reads for the top layer~\cite{2212.11161}
\beq
\label{cont_up}
\partial_t\Phi_\u\cdot\Phi_\u+\partial_xj_{x\u}
=i\frac{\Delta}{\hbar}\Psi^*_\d\sigma_3\Psi_\u-i\frac{\Delta^*}{\hbar}\Psi^*_\u\sigma_3\Psi_\d
=2{\rm sgn}(x)\frac{|\Delta|^2}{v_F\hbar^2}Re(a_1a_2^*)e^{-2|\Delta||x|/\hbar v_F}
\eeq
and for the bottom layer
\beq
\label{cont_down}
\partial_t\Phi_\d\cdot\Phi_\d+\partial_xj_{x\d}
=i\frac{\Delta^*}{\hbar}\Psi^*_\u\sigma_3\Psi_\d-i\frac{\Delta}{\hbar}\Psi^*_\d\sigma_3\Psi_\u
=-2{\rm sgn}(x)\frac{|\Delta|^2}{v_F\hbar^2}Re(a_1a_2^*)e^{-2|\Delta||x|/\hbar v_F}
,
\eeq
since $i(\Delta-\Delta^*)=-2 sgn(x)|\Delta|$. Together with the commutators 
$j_{x}=\frac{i}{\hbar}[H_{\rm EHDL},x]$ and $j_{y}=\frac{i}{\hbar}[H_{\rm EHDL},y]$
we obtain for the current components of the EHDL the same expression as given in 
Eqs. (\ref{qp_current_x}) - (\ref{qp_current_y}).
This agreement of the currents is a consequence of the duality relation between the EEDL and the EHDL.
\begin{figure}[t]
\begin{center}
a)
\includegraphics[width=4cm,height=2.5cm]{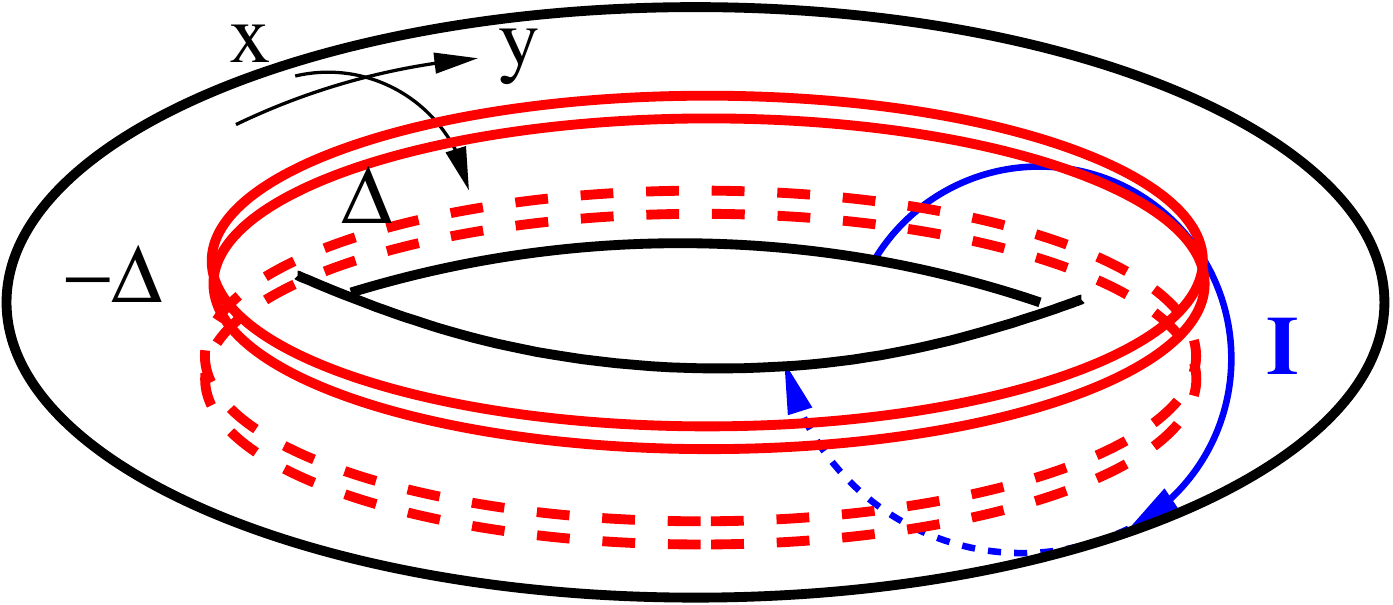}
b)
\includegraphics[width=2cm,height=2.2cm]{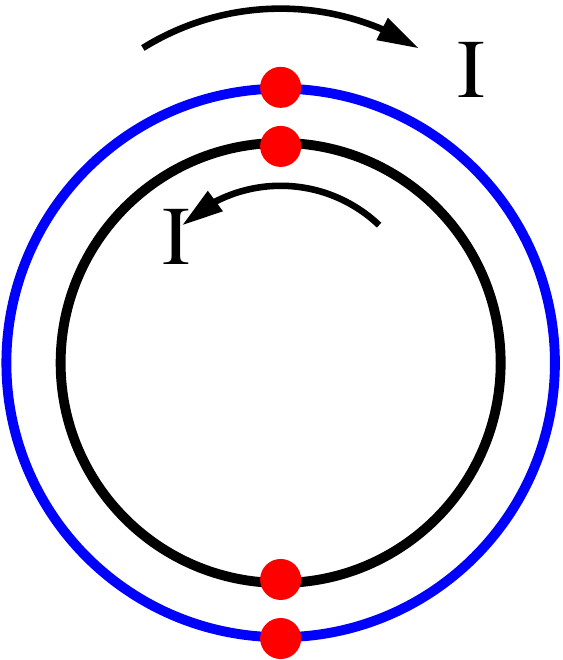}
\caption{
After gluing both layers in Fig. \ref{fig:domain_wall}b) individually, we obtain a double torus with a cross-section 
visualized in b), where each torus has two domain walls.
Then the currents wind along the two domain walls clockwise and counterclockwise around each torus, respectively.
}
\label{fig:torus}
\end{center}
\end{figure}

%%%%%%%%%%%%%%%%%%%%%%%%%%%%%%%%%%%%%%%%%%%%%%%%%
\section{Discussion and Conclusions}
\label{sect:discussion}

The main result of our calculation is the relation between the supercurrent and the zero-energy eigenmodes,
where the latter are characterized by the complex coefficients $a_1$ and $a_2$ in Eqs. (\ref{zero_mode_EEDL})
and (\ref{zero_mode_EHDL}), respectively. This relation reads for the supercurent away from the domain wall as 
\beq
j^s_{x,\u\d}
\sim\mp\frac{ |\Delta|}{\hbar}Re(a_1a_2^*)=\mp\frac{ |\Delta||a_1a_2|}{\hbar}\cos(\varphi_1-\varphi_2)
\ \ \ (|x|\gg \hbar v_F/|\Delta|)
\eeq
for the EEDL as well as for the EHDL, which depends on the phases of the coefficients $a_1$ and $a_2$.
Another interesting result is the angle $\alpha$ between the quasiparticle current and the domain wall
in the top layer 
\beq
\alpha=\arctan (j_{y\u}/j_{x\u})=\frac{\pi}{2}-\varphi_2+\varphi_1
.
\eeq
Thus, the relative phase of the eigenmode coefficients can be tuned by the angle $\alpha$. 

The BdG Hamiltonians $H_{\rm EDDL}$ and $H_{\rm EHDL}$ lead to similar results. In particular, the Josephson
currents are the same for both cases, whereas the zero-energy quasiparticle modes are different.
The origin of this similarity is the duality relation (\ref{duality00}),  (\ref{duality01}) of the two Hamiltonians and
their eigenmodes. Moreover, both Hamiltonians get a sign change under the following transformation
\beq
H_{\rm EEDL}\to TH_{\rm EEDL}T
=- H_{\rm EEDL}
\ ,\ \ 
T=\pmatrix{
 \sigma_3 & 0 \cr
0 & \sigma_3 \cr
}
,
\eeq
and
\beq
H_{\rm EHDL}\to T'H_{\rm EHDL}T'
=- H_{\rm EHDL}
\ ,\ \ 
T'=VTV=\pmatrix{
 \sigma_3 & 0 \cr
0 & -\sigma_3 \cr
}
,
\eeq
which reflects the chirality and the fact that their chiralities are not identical.

All these results indicate that a domain wall or an edge affects the current distribution in the system. Thus, for the general
case we must take into account all edges and domain walls, where the order parameter changes.
On the other hand, we can avoid edges by choosing proper boundary conditions. In $y$ direction we have already
assumed periodic boundary conditions to create a uniform mode in this direction. Assuming two domain
walls (cf. Fig. \ref{fig:domain_wall}b)) and periodic boundary conditions in $x$ direction for both layers
individually, the resulting system is a double torus, which has no edges except for the domain walls, 
as visualized in Fig. \ref{fig:torus}. Then the coefficients $a_1$, $a_2$ are fixed by the matching condition of
the supercurrents in the regions between the domain walls. 

While these considerations give us an idea about the role of the Josephson effect in chiral double layers, 
a complete description requires a solution of the entire microscopic model through a self-consistent approach. 
Then  the supercurrent is induced by an external current or external field, which is represented by a vector potential 
in the BdG Hamiltonian. This external field also affects the order parameter field $\Delta$. In particular, the
creation and measurement of currents in the EHDL was discussed in Ref.~\cite{Su2008}. 

%%%%%%%%%%%%%%%%%%%%%%%%%%%%%%%%%%%%%%%%%%%%%%%%%%
\appendix

\section{Coalescent eigenmodes}
\label{app:exceptional}

For the eigenmodes of the BdG Hamiltonian $H_{\rm EEDL}$ we make the ansatz $\Psi(x)=\psi e^{-bx}$,
where $\psi$ is a four-component spinor and $b$ depends on $x$. This gives for $k_y=0$ with $\bb=v_F\hbar b$ 
the four-dimensional eigenvalue equation
\beq
H^{}_{\rm EEDL}\psi
=\pmatrix{
0 & -i\bb & 0 & -i\Delta\cr
-i\bb & 0 & i\Delta & 0 \cr
0 &-i\Delta & 0 & -i\bb \cr
i\Delta & 0 & -i\bb & 0 \cr
}
\psi=E\psi
\label{H-matrix}
,
\eeq
which has the eigenvalue $E_-=-\sqrt{|\Delta|^2-\bb^2}$ with the pair of eigenspinors
\beq
\label{eigenbasis_eh}
\psi_{1-}=\pmatrix{
1 \cr
0 \cr
\bb/\Delta \cr
-i\sqrt{|\Delta|^2-\bb^2}/\Delta \cr
},\ \ 
\psi_{2-}=\pmatrix{
0 \cr
1 \cr
i\sqrt{|\Delta|^2-\bb^2}/\Delta \cr
-\bb/\Delta \cr
}
\eeq
and the eigenvalue $E_+=\sqrt{|\Delta|^2-\bb^2}$ with the pair of eigenspinors 
\beq
\psi_{1+}=\pmatrix{
1 \cr
0 \cr
\bb/\Delta\cr
i\sqrt{|\Delta|^2-\bb^2}/\Delta  \cr
}
,\ \
\psi_{2+}=\pmatrix{
0 \cr
1 \cr
-i\sqrt{|\Delta|^2-\bb^2}/\Delta \cr
-\bb/\Delta \cr
}
.
\eeq
This indicates a two-fold degeneracy of the eigenvalues $E_\pm$, respectively.
The limit ${\bar b}\to \Delta$ yields $E=0$ and the pairwise coalescent eigenspinors as
\beq
\psi_{1-}\to\psi_{1+}
\to\pmatrix{
1 \cr
0 \cr
1 \cr 
0 \cr 
}
\ ,\ \ 
\psi_{2-}\to\psi_{2+}\to\pmatrix{
0 \cr
1 \cr
0 \cr 
-1 \cr 
}
.
\eeq
Thus, the eigenspace at $E=0$ has only two dimensions, which represents an exceptional point~\cite{Kato:101545}.
This effect is known for solutions of the BdG Hamiltonian with point- and line defects~\cite{Mandal_2015}.

%
%\bibliography{bdg_app}
\end{document}